\documentclass[aps,pra,twocolumn,groupedaddress,floatfix,showpacs]{revtex4}

\usepackage{graphicx}
\usepackage{textcomp}

\begin{document}

\title{~\vspace{2cm}\\
Optical alignment and spinning of laser-trapped microscopic particles}

\author{M. E. J. Friese}

\author{T. A. Nieminen}
\email[]{timo@physics.uq.edu.au}

\author{N. R. Heckenberg}
\author{H. Rubinsztein-Dunlop}

\affiliation{Centre for Laser Science, Department of Physics,
The University of Queensland, Brisbane QLD 4072, Australia}


\begin{abstract}
Light-induced rotation of absorbing microscopic particles by transfer of
angular momentum from light to the material raises the possibility of
optically driven micromachines. The phenomenon has been observed using
elliptically polarized laser beams~\cite{ref1} or beams with helical phase
structure~\cite{ref2,ref3}. But it is difficult to develop high power in such
experiments because of overheating and unwanted axial forces, limiting
the achievable rotation rates to a few hertz. This problem can in
principle be overcome by using transparent particles, transferring
angular momentum by a mechanism first observed by Beth in 1936~\cite{ref4},
when he reported a tiny torque developed in a quartz waveplate due
to the change in polarization of transmitted light. Here we show that
an optical torque can be induced on microscopic birefringent particles
of calcite held by optical tweezers~\cite{ref5}. Depending on the
polarization of the incident beam, the particles either become aligned
with the plane of polarization (and thus can be rotated through specified
angles) or spin with constant rotation frequency. Because these microscopic
particles are transparent, they can be held in three-dimensional optical
traps at very high power without heating. We have observed rotation rates in
excess of 350\,Hz.

\vspace{-9.5cm}
{\small\noindent
\textbf{Preprint of:}\\
M. E. J. Friese, T. A. Nieminen, N. R. Heckenberg and
H. Rubinsztein-Dunlop\\
``Optical alignment and spinning of laser-trapped microscopic particles''\\
\textit{Nature} \textbf{394}, 348--350 (1998)\\
erratum in \textit{Nature} \textbf{395}, 621 (1998)
}
\vspace{8.2cm}

\end{abstract}


\maketitle

A typical optical tweezers arrangement was used to trap microscopic
calcite particles in three dimensions using between 30 and 300\,mW
of laser light at a wavelength of 1064\,nm. The optical trap used a
$100\times$ oil-immersion, high numerical aperture (NA $ = 1.3$)
microscope objective. The trapping beam was initially linearly polarized,
and the plane of polarization could be rotated using a half-wave plate.
Alternatively, a quarter-wave plate allowed the ellipticity of
polarization to be varied. The particles were fragments obtained
by crushing a small calcite crystal, giving irregular particles
1--15\,{\textmu}m across. They were dispersed in distilled water
in a trapping cell consisting of a well in a microscope slide with
a coverslip.

Because of their birefringent nature, calcite particles can act as
wave-plates; a calcite particle 3\,{\textmu}m thick is a $\lambda/2$ plate
for 1064\,nm light. On passage through a fragment of calcite, the
ordinary and extraordinary components of the incident light will
undergo different phase shifts. If this results in a change in the
angular momentum carried by the light, there will be a corresponding
torque on the material. Our results can be understood using a
simple plane wave picture; the interaction between an incident plane
wave and a waveplate is outlined below. We note that the calcite
waveplate is trapped at the focal point of the beam, where the wavefronts
are nearly plane.

An incident laser beam can, in general, have both circularly polarized
and plane polarized components; that is, it will be elliptically
polarized. Elliptically polarized light can be described by
\begin{displaymath}
\mathbf{E} = E_0 \exp(-\mathrm{i}\omega t) \cos\phi \hat\mathbf{x}
+ \mathrm{i} E_0 \exp(-\mathrm{i}\omega t) \sin\phi \hat\mathbf{y}
\end{displaymath}
where $\phi$ describes the degree of ellipticity of the light
($\phi = 0$ or $\pi/2$) indicates plane-polarized light, $\phi = \pi/4$
circularly polarized light). The angular momentum of a plane
electromagnetic wave (the incident light) of angular frequency $\omega$
can be found from the electric field $\mathbf{E}$ and its complex conjugate
$\mathbf{E}^\star$ by integrating over all spatial elements $\mathrm{d}^3 r$
giving
\begin{displaymath}
\mathbf{J} = \frac{\epsilon}{2\mathrm{i}\omega} \int \mathrm{d}^3 r
\mathbf{E}^\star \times \mathbf{E},
\end{displaymath}
where $\epsilon$ is the permittivity.

To calculate the change in angular momentum of the light after passage
through a birefringent material, the incident elliptically polarized
light is first expressed in terms of components parallel and perpendicular
to the optic axis of the material by
\begin{eqnarray}
\mathbf{E} & = & E_0 \exp(-\mathrm{i}\omega t)
(\cos\phi \cos\theta - \mathrm{i} \sin\phi \sin\theta ) \hat\mathbf{i}
\nonumber \\
& & + E_0 \exp(-\mathrm{i}\omega t)
(\cos\phi \sin\theta + \mathrm{i} \sin\phi \cos\theta ) \hat\mathbf{j}
\end{eqnarray}
where $\theta$ is the angle between the fast axis of the quarter-wave
plate producing the elliptically polarized light and the optic axis of
the birefringent material. The phase shift due to passing through a
thickness $d$ with refractive index $n$ is $kdn$, where $k$ is the
free-space wavenumber, so the emergent light field will be
\begin{eqnarray}
\mathbf{E} & = & E_0 \exp(-\mathrm{i}\omega t) \exp(\mathrm{i}kdn_e)
\nonumber \\ & &
(\cos\phi \cos\theta - \mathrm{i} \sin\phi \sin\theta ) \hat\mathbf{i}
\nonumber \\
& & + E_0 \exp(-\mathrm{i}\omega t) \exp(\mathrm{i}kdn_o)
\nonumber \\ & &
(\cos\phi \sin\theta + \mathrm{i} \sin\phi \cos\theta ) \hat\mathbf{j}
\end{eqnarray}
where $n_e$ and $n_o$ are the extraordinary and ordinary refractive indices
of the birefringent material.

\begin{figure}[b]
\includegraphics[width=\columnwidth]{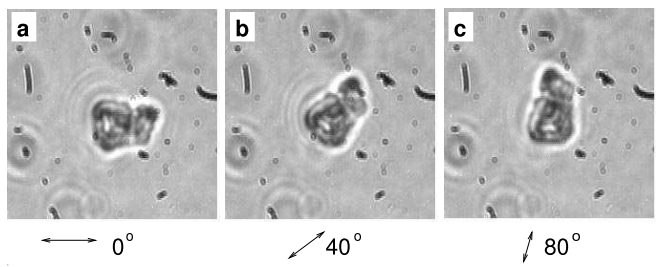}%
\caption{Three sequential photographs (frames) of a trapped calcite crystal,
showing alignment with the plane of polarization of the trapping beam.
A $\lambda/2$ waveplate was rotated by 20\textdegree between successive
photographs, rotating the plane of
polarization by 40\textdegree, as shown by the arrows, and
exerting an alignment torque on the crystal, causing it to rotate
to a new position. This can be used to rotate the particle at a
controlled speed, or to control its orientation.
\label{fig1}}
\end{figure}

The changes in the angular momentum of the light cause a reaction torque
per unit area on the thickness $d$ of material of
\begin{eqnarray}
\tau & = & -\frac{c\epsilon}{2\omega} E_0^2 \sin( kd(n_0 - n_e))
\cos 2\phi \sin 2\theta
\nonumber \\
& & + \frac{c\epsilon}{2\omega} E_0^2 \{ 1 - \cos( kd(n_0 - n_e))\} \sin 2\phi
\label{eqn3}
\end{eqnarray}
In general, the first term of equation (\ref{eqn3}) is the torque due to the
plane-polarized component of elliptically polarized light while the
second term is due to the change in polarization caused by passage
through the medium. For plane-polarized light, $\phi = 0$ or $\pi/2$,
so the torque on the particle is proportional to $\sin 2\theta$,
so that a particle will experience torque so long as $\theta$ is non-zero,
and will be at a stable equilibrium when the fast axis of the crystal is aligned
with the plane of polarization ($\theta = 0$). We found that calcite
fragments trapped in plane-polarized light are aligned in a particular
orientation, and a particular particle is always aligned in the same
plane each time it is trapped. When the plane of polarization is rotated
using a half-wave plate, a particle's alignment exactly follows the
rotation of the plane of polarization. In figure~\ref{fig1}, a calcite
fragment is shown to rotate through 80\textdegree\ as a half-wave plate
controlling the polarization of the trapping beam is rotated through
40\textdegree, illustrating the alignment of birefringent particles
to the plane of polarization. To our knowledge,
this is the first report of an optically trapped particle being
rotated through a preset angle; a modification to the setup whereby
the half-wave plate could be spun at a set rate would also allow the
particle to rotate at a preset frequency.

\begin{figure}[t]
\includegraphics[width=\columnwidth]{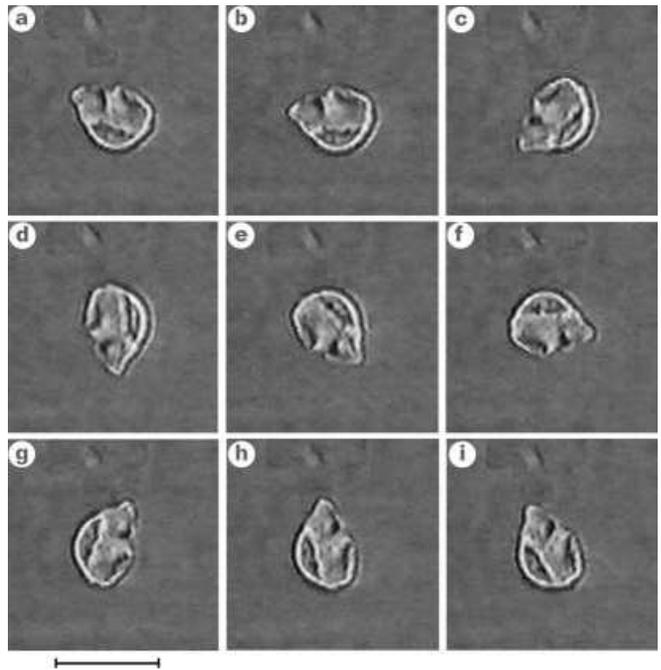}%
\caption{Nine frames of a trapped calcite crystal, showing free rotation due to an
elliptically polarized trapping beam. The speed of rotation is limited by
the viscous drag on the particle. As the optical torque acting on the
particle depends on its orientation, the rotation speed is not constant.
The frames are 40\,ms apart. Scale bar is 10\,{\textmu}m.
\label{fig2}}
\end{figure}

The second term of equation (\ref{eqn3}) will be constant for a given
laser power and ellipticity of polarization (characterized by $\phi$),
and will be maximum for circularly polarized light when the first
term vanishes. Hence, when trapped in a circularly polarized beam,
a birefringent particle will experience constant torque. In a viscous
medium, this torque will be balanced by the drag torque, $t_D = D\Omega$,
where $D$ is the drag coefficient and $\Omega$ is the angular speed,
so in this case a birefringent particle will rotate with constant
frequency and angular speed. We measured the rotation frequencies of trapped
calcite fragments by detection of back-scattered light~\cite{ref1};
the results show that calcite fragments rotate at constant frequency
in circularly polarized light, and that this frequency is proportional
to the laser power. A rotating particle is shown in figure \ref{fig2}.
The fastest rotation frequency measured was 357\,Hz, for a particle
1\,{\textmu}m thick trapped in a 300\,mW laser beam.

\begin{figure}[t]
\includegraphics[width=\columnwidth]{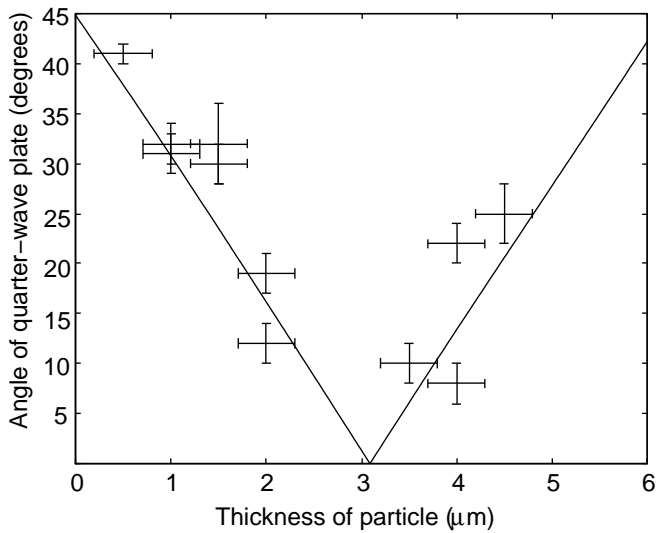}%
\caption{The degree of circular polarization required to cause spinning of a
trapped particle depends on its thickness. Here we compare
measurements of the minimum angle required for rotation with the
theoretical solution. Only if the particle is the exact thickness
of a half-wave plate will it always spin in elliptical light.
For all other particle thicknesses, there is some angle $\theta$
for which the maximum alignment torque will be greater than
the spinning torque. The degree of ellipticity of polarization
(measured by $\phi$) required for the onset of rotation is found
from the case where the alignment torque is maximum and the total
torque is zero, which is when
$\sin[kd(n_o - n_e)]\cos 2\phi = \{ 1 - \cos[kd(n_o - n_e)]\} \sin 2\phi$.
The solution to this is
$\phi_{\mathrm{rotate}} = [\pi - kd(n_o - n_e)]/4$.
In general a particle will be aligned to the plane of polarization
of the trapping beam unless there is sufficient torque due to
the circularly polarized component to set it into rotation,
when it will experience a position-dependent torque.
\label{fig3}}
\end{figure}

\begin{figure}[t]
\includegraphics[width=\columnwidth]{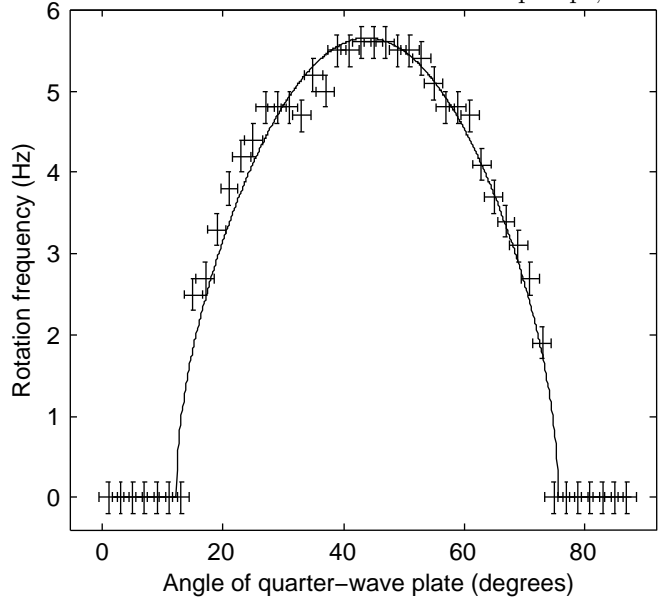}%
\caption{The variation of rotation frequency with the polarization of the
trapping beam. The sudden onset of rotation when the minimum angle
is reached can be clearly seen. The theoretical response of the particle
allows a particle of unknown size to be measured, or can be used to
determine the viscosity of a fluid. In this case, the trapping beam has
a power of 50\,mW, and the particle is 6.15\,{\textmu}m in radius and
2.3\,{\textmu}m thick. The frequency of rotation for a trapping beam of
power $P$ and optical frequency $\omega$ is
$f = [\mathrm{Re}(P\{[1 - \cos kd(n_o - n_e)]^2 \sin^2 2\phi
- \sin^2 kd(n_o - n_e) \cos^2 2\phi\}^{1/2})]/(2\pi\omega D)$.
The drag torque coefficient $D$ can be estimated by representing the
particle as an ellipsoid~\cite{ref6}. The drag will lie between that
of a sphere of radius $a$ in a medium of viscosity
$\mu$ ($D = 8\pi\mu a^3$) and a disk of the same radius
($D = (32/3)\mu a^3$). Under the very low Reynolds number flow
conditions encountered here, the surface texture and fine structure
of the particle are unimportant. The maximum rotation speed,
$f = P/\pi\omega D$, will result when the incident light is
circularly polarized and the particle is a half-wave plate.
Small particles will generally rotate faster due to less drag,
but as particles become too small, their thickness becomes much
less than the ideal half-wave case, and they will not intercept
all of the power available to spin larger particles.
\label{fig4}}
\end{figure}

For the general case of elliptically polarized light, both the alignment
torque and the spinning torque will act, and the effect on the
birefringent particle will depend on the thickness $d$ and the ellipticity
$\phi$ of the light. The particle will only rotate if the maximum
alignment torque is less than the spinning torque. This is shown in
figure \ref{fig3}. In figure \ref{fig4} we plot the variation of
rotation rate of a larger calcite crystal with degree of ellipticity
of the trapping beam $\phi$, showing the characteristic behaviour of a
birefringent particle in elliptically polarized light.

The agreement between our results and the theory outlined above shows that
the calcite particles act as microscopic wave-plates. The measurement of
the rotation speed of spinning particles is a less accurate but simpler
analogue of Beth's experiment~\cite{ref4}. In particular, assuming
conventional viscous drag, the observed speeds are consistent with
the accepted intrinsic spin angular momentum of $\hbar$ per photon
(see figure \ref{fig4}). Our results also show how optical torques can
be exerted on certain microscopic objects with high precision and
efficiency, with minimal heating. Depending on the details of the
arrangement, either a constant torque independent of orientation can be exerted,
leading to rotation rates up to hundreds of hertz, or alternatively, the
orientation of the object can be smoothly controlled.

The controllability of the motion of calcite particles and the
minimal absorption involved suggests calcite as an ideal material for
optically driven rotary micromachines. Such micromachines could include
pumps, stirrers, or optically powered cogwheels. The rotation could also
be used to study the viscosity of small samples of fluids, or the alignment
of calcite particles could be used to hold probe particles in particular
orientations, which could be useful for atomic-force or other forms of
microscopy. The birefringence of biological samples is usually much less
than that of calcite, but may sometimes be large enough to allow the same
alignment and free rotation to be achieved.

\end{document}